\documentclass[prl,aps,twocolumn,a4paper,showpacs,superscriptaddress,10pt]{revtex4}
\usepackage{amsmath,amssymb}
\usepackage{graphicx}
\usepackage{color}
\usepackage{hyperref}
\usepackage[cm]{fullpage}

\newcommand{\mytitle}{Strong-Field Many-Body Physics and the Giant Enhancement in the High-Harmonic Spectrum of Xenon}

\newcommand{\rmpdfinfo}{\special{ps:: userdict /pdfmark /cleartomark load put}}

\definecolor{MyDarkGreen}{rgb}{0,0.6,0}
\definecolor{MyDarkBlue}{rgb}{0,0,0.8}
\definecolor{MyDarkRed}{rgb}{0.6,0,0.3}
\hypersetup{breaklinks=true, colorlinks=true,plainpages=true, linktocpage=true, linkcolor=MyDarkBlue, citecolor=MyDarkGreen, urlcolor=MyDarkRed, pdfborder={0 0 0},%
pdfauthor={Stefan Pabst},%
pdfsubject={Research article},
pdftitle={\mytitle}%
}

\newcommand{\ket}[1]{\left|#1\right>}
\newcommand{\bra}[1]{\left<#1\right|}

\begin{document}

\title{\mytitle} 

\author{Stefan Pabst}
\email[]{stefan.pabst@cfel.de}
\affiliation{Center for Free-Electron Laser Science, DESY, Notkestrasse 85, 22607 Hamburg, Germany}

\author{Robin Santra}
\email[]{robin.santra@cfel.de}
\affiliation{Center for Free-Electron Laser Science, DESY, Notkestrasse 85, 22607 Hamburg, Germany}
\affiliation{Department of Physics,University of Hamburg, Jungiusstrasse 9, 20355 Hamburg, Germany}

\date{\today}


\begin{abstract}
We resolve an open question about the origin of the giant enhancement in the high-harmonic generation (HHG) spectrum of atomic xenon around 100~eV.
By solving the many-body time-dependent Schr\"odinger equation with all orbitals in the $4d, 5s$, and $5p$ shells active, we demonstrate the enhancement results truly from collective many-body excitation induced by the returning photoelectron via two-body interchannel interactions.
Without the many-body interactions, which promote a $4d$ electron into the $5p$ vacancy created by strong-field ionization, no collective excitation and no enhancement in the HHG spectrum exist.
\end{abstract}

\maketitle


High-harmonic generation (HHG) is one of the most fundamental processes in attosecond physics~\cite{KrIv-RMP-2009,KoPf-AAMO-2012,Pa-EPJST-2013}.
Initially HHG was primarily used to generate attosecond light pulses~\cite{AgDi-RPP-2004} by converting light from the near-infrared (NIR) into the ultraviolet (UV)~\cite{PaCh-PRL-1999,PaTo-Science-2001} and x-ray regime~\cite{PoCh-Science-2012}. 
The shortest pulse generated is as short as 67~attoseconds~\cite{ZhZh-OptLett-2012}.
Combined with the NIR driving pulses, pump-probe experiments~\cite{GoKr-Nature-2010,WiGo-Science-2011,OtKa-Science-2013,DrKr-Nature-2002,ItQu-PRL-2002,ScYa-Science-2010} with a time delay control of a few attoseconds~\cite{WiGo-Science-2011} and below~\cite{KoWo-OptExp-2011} can be performed.

More recently, the HHG process has also been used as a probe mechanism to study electronic structure~\cite{ItLe-Nature432-2004,McGu-Science-2008,SmIv-Nature-2009,HaCa-JPhysB-2011} and electron dynamics~\cite{ShSo-Nature-2012,ShTi-PRL-2012}. 
A prominent example is orbital tomography~\cite{ItLe-Nature432-2004} where the form factor of the outer-most orbitals is encoded in the HHG spectrum.
Therefore, it is of high relevance to understand how the electronic structure imprints itself on the HHG spectrum.
In atomic systems, several characteristic phenomena have been observed in the HHG spectrum ranging from the Cooper minimum near 50~eV in argon~\cite{WoVi-PRL-2009,HiFa-PRA-2011,FaSc-PRA-2011,PaGr-PRA-2012} to the giant resonance enhancement in xenon around 100~eV~\cite{ShVi-NatPhys-2011}.

The mechanism behind the enhancement in the xenon HHG spectrum, which has been theoretically~\cite{ZhGu-PRL-2013} and experimentally~\cite{ShVi-NatPhys-2011} studied, is still not understood.
Shiner {\it et al.}~\cite{ShVi-NatPhys-2011} argued that the enhancement originates from a collective resonance involving the $4d$ electrons, which are much tighter bound (at 68~eV) than are the outer-valence $5p$ electrons (with a binding energy of 12~eV).
Zhang and Guo~\cite{ZhGu-PRL-2013} questioned this interpretation and attributed the origin of the enhancement to the Cooper minimum of the $5p$ shell, which can be fully explained with a single-electron picture.
Furthermore, additional studies~\cite{VoNe-NJP-2011} have shown that the existence of the enhancement starting around 60-70~eV depends on the geometric Gouy-phase and can even vanish for certain values.
Does this suggest the enhancement might be a propagation and not a single-atom response effect?

In this letter, we answer this open question and show that the enhancement is a single-atom response effect.
We show that the giant enhancement in the HHG spectrum is truly a collective many-body effect involving the $4d$ electrons via the two-body electron-electron interaction.
Our results are based on solving the many-body Schr\"odinger equation for atomic xenon with 18 active electrons (6 in the $5p$ shell, 2 in the $5s$ shell, and 10 in $4d$ shell) in the presence of a strong-laser field with the time-dependent configuration-interaction singles (TDCIS) approach.
The enhancement disappears when many-body interactions with the $4d$ orbitals are excluded from our calculations and when the photoelectron is allowed to recombine only with the $5p$ orbitals.

The particularly large $4d$ recombination matrix elements (in comparison with the $5p$ matrix elements) leads to an order-of-magnitude enhancement in the HHG spectrum.
We show that all five $4d_m$ orbitals ($m=0, \pm 1, \pm 2$) contribute to the HHG spectrum and it is not sufficient to focus only on the orbital aligned in the direction of the laser polarization (i.e., only $4d_0$).
When done so the resulting HHG enhancement region would be 20~eV too low with only a 15~eV spectral width.

So far, other time-dependent many-body approaches in the strong-field regime are not able to perform calculations on xenon with NIR wavelengths of around $1.5~\mu$m. 
Most many-body approaches focus on two-electron model systems~\cite{SuCh-PRA-2002,Sc-NJP-2012,MiMa-PRA-2013}, helium~\cite{ScPi-PRA-1998,PrHu-PRA-2001,PaMe-JPB-2007,DjSt-PRA-2011,TaGr-PRA-2012,HoBo-PRA-2012}, berylium~\cite{HoBo-PRA-2012} or H$_2$~\cite{SaPa-JESRP-2007}.
The application to larger systems with more than two electrons in the outer-most shell (like most noble gas atoms) is challenging.
The time-dependent R-matrix approach has already been applied to larger systems like argon~\cite{BrRo-PRA-2012,BrHu-PRL-2012} with relative short wavelengths (up to 390~nm)~\cite{n.hugo}.

The challenge in the extension to long-wavelength driver is two-fold.
First, the spatial extend of the electron motion in the continuum increases with $\lambda^2$.
Second, the kinetic energy the electron gains in the strong-field increases also with $\lambda^2$.
Taking both aspects together, we find that the size of the numerical grid grows with $\lambda^4$.
This results in a large one-particle basis which strongly limits the complexity of correlation effects that can be captured in these long-wavelength-driven many-electron systems.
In most HHG calculations, the single-active electron (SAE) approximation is used where a one-electron problem is solved without any kind of correlation effects~\cite{HiFa-PRA-2011}.

Our many-body TDCIS approach~\cite{n.xcid} has been described in detail in previous publications~\cite{GrSa-PRA-2010,PaGr-PRA-2012} and has been successfully applied to perturbative~\cite{PaSa-PRL-2011,SyPa-PRA-2012} and non-perturbative~\cite{PaGr-PRA-2012,PaSy-PRA-2012,KaPa-PRA-2013} multiphoton processes with photon energies from the x-ray down to the NIR regime.
The TDCIS wave function ansatz reads

\begin{align}
  \label{eq1}
  \ket{\Psi(t)}
  =
  \alpha_0(t) \, \ket{\Phi_0}
  +
  \sum_{a_i}
    \alpha^a_i(t) \, \ket{\Phi^a_i}
  ,
\end{align}
where $\ket{\Phi_0}$ is the Hartree-Fock ground state and $\ket{\Phi^a_i}= \hat c^\dagger_a \hat c_i \ket{\Phi_0}$ are the singly excited configurations with one electron removed from the initially occupied orbital $i$ and placed in the initially unoccupied orbital $a$.
The resulting equations of motion for the CIS coefficients are given by:

\begin{subequations}
\label{eq2}
\begin{align}
  \label{eq2.1}
  i\partial_t \, \alpha_0(t)
  &=
  -E(t)\, \sum_{a,i} \bra{\Phi_0} \hat z \ket{\Phi^a_i}
  \\\nonumber
  \label{eq2.2}
  i\partial_t \, \alpha^a_i(t)
  &= 
  \bra{\Phi^a_i} \hat H_0 \ket{\Phi^a_i} \, \alpha^a_i(t)
  +
  \sum_{b,j}
    \bra{\Phi^a_i} \hat H_1 \ket{\Phi^b_j}
    \alpha^b_j(t)
  \\ &
  -
  E(t)\, \Big(    
    \alpha_0(t)
    \bra{\Phi^a_i} \hat z \ket{\Phi_0}
    \! + \!
    \sum_{jb} \bra{\Phi^a_i} \hat z \ket{\Phi^b_j}
    \alpha^b_j(t)
    \!
  \Big)
  ,
\end{align}
\end{subequations}
where $\hat H_0$ includes all one-particle operators including the mean-field potential $\hat V_\text{MF}$, and a constant energy shift by $-E_\text{HF}$, where $E_\text{HF}$ is the Hartree-Fock
energy, such that the Hartree-Fock ground state, $\ket{\Phi_0}$, has the energy 0~a.u.
The light-matter interaction for linearly polarized pulses is given by $-E(t)\, \hat z$, and all electron-electron interactions that cannot be described by the one-particle mean-field potential $\hat V_\text{MF}$~\cite{RoSa-PRA-2006} are captured by $\hat H_1 = 1/2\sum_{ij} 1/|\hat{\bf r}_i - \hat{\bf r}_j| - \hat V_\text{MF}$.
The TDCIS approach includes many-body physics that goes beyond an independent particle picture~\cite{SzOs-book,KrPa-submitted}.
Specifically, interchannel interactions are included where the excited/ionized electron changes the state of the parent ion by exchanging energy.
We will see that exactly these interchannel interactions lead to the enhancement in the HHG spectrum

To reveal the collective resonance in xenon, centered around 100~eV, we have chosen an HHG cut-off energy of 157~eV, which is reached by a strong-field NIR pulse with a driving wavelength of $1.69~\mu$m, a pulse duration of 10~fs at FWHM, and a peak intensity of $1.7 \times 10^{14}$~W/cm$^2$.
Numerically this requires a box size of $250\,a_0$ and a maximum angular momentum of $L=100$ for describing the field-driven electron in the continuum (more parameters can be found in Ref.~\cite{n.parameters}).
The description of the electron-electron interaction, which leads to the many-body correlation effects, is especially challenging due to the size of the Hamiltonian matrix ($260,000 \times 260,000$) and the numerical effort to calculate its action on the wave function [see Eq.~\eqref{eq2}].

\begin{figure}[t!]
  \centering
  \includegraphics[clip,width=\linewidth]{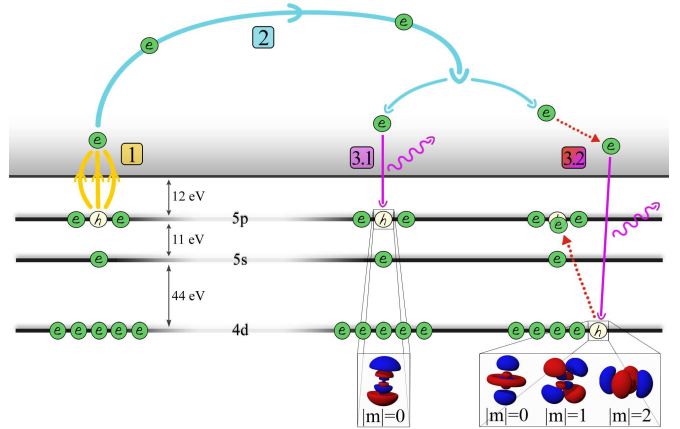}
  \caption{(color online) Schematic illustration of the HHG process in xenon (spin states are excluded).
    (1) The electron is tunnel-ionized mainly from the $5p_0$ orbital by the strong-field driving pulse,.
    (2) The electron is driven back to the ion by the oscillating electric field.
    In the third step the electron recombines with the ion in two different ways: 
    (3.1) the electron recombines with the very same hole that was generated in step 1,
    or (3.2) the electron exchanges energy with the ion by promoting an inner shell electron from the $4d$ shell in the $5p_0$ hole via Coulomb interaction before recombining in a more tightly bound $4d$ orbital.
    All five $4d_m$ orbitals ($m=-2,\ldots,2$) contribute to the Coulomb interaction.    
  }
  \label{f1}
\end{figure}

A schematic of the HHG process for atomic xenon is shown in Fig.~\ref{f1}.
The first two steps are the same as in the well-known 3-step model~\cite{Co-PRL-1993,ScKu-PRL-1993,LeCo-PRA-1994}.
In step 1, the $5p_0$ electron is predominantly tunnel ionized by the NIR strong-field pulse, and in step 2, the electron moves in the continuum driven by the strong-field pulse.
The third step, where the electron recombines with the parent ion, can happen in two ways:
either (3.1) the electron recombines as usual with the $5p_0$ orbital, or (3.2) the returning electron and the ion exchange energy via the electron-electron interaction just before recombination so that the electron recombines with an excited ionic state where the hole is instead in the $4d$ shell.

The electron-electron Coulomb interaction mediating this exchange of energy can also change the magnetic $m$ state of the electrons involved.
Therefore, all five $4d_m$ orbitals with $m=-2,\ldots,2$ are equally important (as is seen in the later discussion). 
The two-body character of the electron-electron interaction involved in step 3.2 exists only in many-body but not in single-electron theories.

Within the TDCIS theory, the interchannel interaction responsible for step 3.2 can be turned off or on for each calculation.
In this way, the electron is allowed to recombine either only via step 3.1 or via both steps (3.1 and 3.2), respectively.
Besides the physical mechanisms, the active orbitals participating in the HHG process can also be controlled.
Here, we consider in particular three different scenarios:
\begin{itemize}
\item 
all 9 orbitals~\cite{n.spin} in the $4d, 5s,$ and $5p$ shells are active and all interchannel and intrachannel interactions (included in TDCIS) are allowed,
\item 
only the orbitals aligned with the laser polarization ($4d_0, 5s, 5p_0$) are active and all interchannel and intrachannel interactions are allowed,
\item all 9 orbitals in the $4d, 5s,$ and $5p$ shells are active and only intrachannel interactions are allowed.
\end{itemize}
Even though the first two scenarios are artificially simplified by ignoring certain interactions are orbitals, they are of high educational purpose since the direct comparison with the most complete third scenario will enable us to identify the underlying mechanisms.  

\begin{figure}[t!]
  \centering
  \rmpdfinfo
  \includegraphics[clip,width=\linewidth]{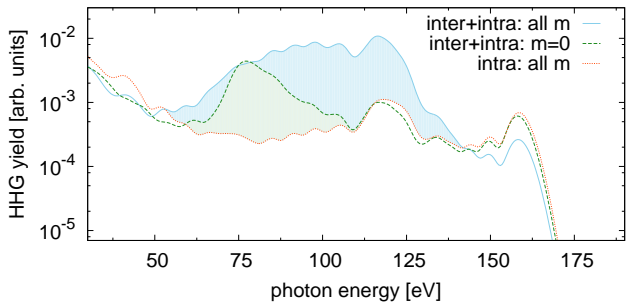}
  \caption{(color online)  
    The HHG spectrum of xenon for different theoretical models:
    (solid blue) the full TDCIS model (intrachannel + interchannel interactions) with all orbitals in the $4d, 5s,$ and $5p$ shells active,
    (dashed green) the full TDCIS model with only $m=0$ electrons in the $4d, 5s,$ and $5p$ shells active, and
    (dotted red) the TDCIS model, excluding interchannel coupling, with all orbitals in the $4d, 5s,$ and $5p$ shells active.
    The difference between the two models including interchannel effects and the model including only intrachannel model is highlighted with the according colors.
  }
  \label{f2}
\end{figure}

Figure~\ref{f2} shows the HHG spectra~\cite{n.velocity} for these three cases.
When all 9 orbitals are active and only intrachannel interactions are included (dotted red line), the HHG yield has the typical form of an atomic HHG spectrum with a flat plateau region up to the cut-off energy---no enhancement can be seen.
This results does not change when only the $5p$ shell is considered (not shown) indicating that the direct contributions from $4d$ and $5s$ are negligible. 

By including the many-body interchannel interactions in the calculations, a strong enhancement of up to one order of magnitude in the HHG yield appears (solid blue and dashed green lines).
Depending on which orbitals are active the width and center of the enhancement region are quite different. 
When only laser-aligned $4d_0, 5s$, and $5p_0$ orbitals are active, the enhancement is located in the energy region of 60-90~eV compared with an enhancement ranging from 60 to 125~eV when all electrons in the $4d, 5s$, and $5p$ shells are active.

The effect of interchannel coupling is dominated by the coupling between the $5p_0$ hole created in step 1 and the $4d_m$ orbitals. 
Unlike the interaction with the light field, where the magnetic quantum number $m$ of each electron cannot change, the electron-electron interaction can change the $m$ state.
Only the overall quantum number $M=\sum_p^N m_p$ of the entire $N$-electron system is conserved but the $m_p$ for the individual electron $p$ is not conserved.
Considering only the $4d_0$ orbital is not sufficient as this leads to enhancements which are much too narrow (as shown in Fig.~\ref{f2}).

The position and the width of the giant enhancement in the HHG spectrum (see Fig.~\ref{f2}) is not arbitrary and can be explained by examining the photoionization cross sections (PICS).
The close connection between cross sections and the recombination cross sections, and therefore to the HHG spectra, has been established by the quantitative rescattering theory~\cite{MoLi-PRL-2008,LeLi-PRA-2009} and successfully applied to various systems.
One aspect that has not been explicitly discussed is whether the partial or total PICS should be used to establish this connection.
Especially in the case where returning electron is highly energetic and can induce ionic excitations such that more than one occupied orbital is involved, it is not clear whether the partial or the total PICS should be considered.

\begin{figure}[t!]
  \centering
  \rmpdfinfo
  \includegraphics[clip,width=\linewidth]{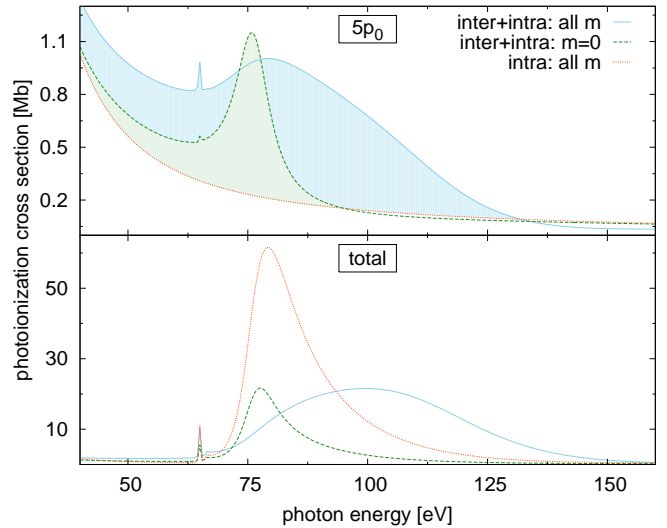}
  \caption{(color online) 
    The partial photoionization cross sections of the $5p_0$ orbitals and the total photoionization cross section of all orbitals for the three different theoretical models (see Fig.~\ref{f2}).    
  }
  \label{f3}
\end{figure}

In the following, we show it is always the partial PICS that should be used~\footnote{
In the case the HHG process involves only one orbital at all times (as in a single-active electron approximation), the use of the PICS has been already stated in Ref.~\cite{MoLi-PRL-2008}.
}. 
Figure~\ref{f3} shows the partial $5p_0$ (a) and the total (b) photoionization cross sections~\cite{n.velocity} of atomic xenon for all three CIS models.
By comparing the cross sections for each theoretical model with its corresponding HHG spectrum (shown in Fig.~\ref{f2}), one clearly sees that the strong enhancement in the HHG yield is, indeed, also present in the photoionization cross section.
A careful comparison reveals that only the partial $5p_0$ PICS (see Fig.~\ref{f3}a) can explain the HHG spectrum of all three CIS models.
Also when considering the recombination step, one finds that the initial conditions of the ionic state (having dominantly a $5p_0$ hole) do only coincide with the partial and not the total PICS.
tunnel process (step 1 in Fig.~\ref{f1})
Particularly when no interchannel interactions are allowed, the total cross section still shows an enhancement due to the continuum resonance of the direct $4d$ ionization channel~\cite{St-Springer-1980}, which is totally absent in the HHG spectrum.

For the total PICS, the ion is produced in any energetically allowed state. 
The inverse process assumes therefore that the ion has up on recombination a hole in any of these accessible orbitals.
This is however not the case due to tunnel ionization and all orbitals that are not accessed via tunnel ionization can only contribute indirectly to the HHG process but never directly (what stands in contradiction with the use of the total PICS).



Lastly, we consider the electron spectrum of the returning electron $W(\omega)$.
It is commonly used that the spectrum of the returning electron is universal and not system-dependent~\cite{MoLi-PRL-2008,ShVi-NatPhys-2011}.
This sounds especially reasonable for atomic systems because for already small distances~\cite{n.radius}, $r \gtrsim 5~a_0$, the electron 'experiences' the parent ion as a singly-charged spherically-symmetric object making the dynamics of the electron in step 2 (see Fig.~\ref{f1}) system-independent.
However, recombination happens very close to the atom~\cite{n.radius}, $r \lesssim 3~a_0$, where many-body effects like interchannel interactions can no longer be neglected. 
This short-range potential is highly system- and state-dependent and it is very much likely that it also affects the spectrum of the returning electron.

\begin{figure}[t!]
  \centering
  \rmpdfinfo
  \includegraphics[clip,width=\linewidth]{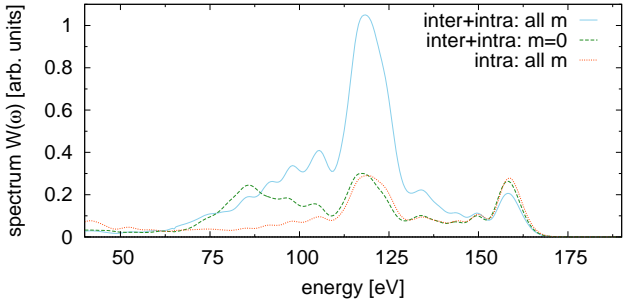}
  \caption{(color online) 
    The returning electron spectrum $W(\omega)$ calculated by using the factorization 
    $W(\omega) = S(\omega) / \sigma(\omega)$,
    where $S(\omega)$ is the HHG spectrum and $\sigma(\omega)$ is the photoionization cross section.
    The electron spectra shown for the three different theoretical models (see Fig.~\ref{f2}) illustrate the system-dependence of the returning electron wavepacket---especially when interchannel interactions are not negligible.
  }
  \label{f4}
\end{figure}

We study this approximation by comparing the spectrum of the returning electron for the three different CIS models (see Fig.~\ref{f4}).
Each CIS model describes different electron-electron interactions and consequently different short-range potentials.
For large distances, all three models turn into a long-range Coulomb potential, $-1/r$.

The three electron spectra, $W(\omega) = S(\omega) / \sigma(\omega)$~\cite{MoLi-PRL-2008,n.espec}, shown in Fig.~\ref{f4} agree well with each other above and below the giant resonance of the $4d$ shell, where $S(\omega)$ is the HHG spectrum and $\sigma(\omega)$ is the partial $5p_0$ PICS of the corresponding models.
In the region of the giant enhancement ($60-140$~eV) the spectra start to differ from each other and reaches a maximum around 120~eV.
The enhancement seen in the HHG spectra and in the cross sections persists also in the photoelectron spectrum indicating that the enhancement in the HHG yield cannot be solely explained by the modified cross sections.
It demonstrates that the modified electron-electron interaction influences not just the recombination matrix elements but also the electron spectrum, which in the region of $r \lesssim 3~a_0$ is sensitive to the modified short-range potential.
Note, however, that the electron spectra in Fig.~\ref{f4} are plotted in a linear scale and the differences in the electron spectra are smaller than in the HHG spectra, which are plotted logarithmically (see Fig.~\ref{f2}). 

In conclusion, we performed ab-initio many-body calculations based on the time-dependent configuration interaction singles (TDCIS) approach in the strong-field regime, which is applicable to closed-shell atoms as large as xenon.
By limiting the classes of electronic excitations to singles we are able to describe up to 18 active electrons and the interactions among them.

We identified the many-body interchannel interactions between the $5p$ orbitals and all five $4d$ orbitals to be the underlying mechanism behind the giant enhancement in the HHG spectrum of atomic xenon. 
This collective excitation happens only when the returning electron is close by and can exchange energy. 
Limiting the electron-electron interaction to the aligned ionic states $5p_0^{-1}$ and $4d_0^{-1}$ (as it is usually done in step 1 of the HHG process) is insufficient and leads to quantitatively incorrect enhancements.
As demonstrated, the details of the short-range potential can also affect the returning electronic wave packet and, consequently, the spectrum of the recombining electron.
This seems to be especially the case when many-body effects start to kick in.
To understand the detailed structure of the HHG spectrum, all aspects of the many-body effects have to be considered, which (as it seems) cannot be fully summarized by a corrected cross section.

Our results demonstrate the need for many-body theories in the strong-field regime.
For molecular systems multi-orbital and many-body effects are even more prominent~\cite{McGu-Science-2008,SmIv-Nature-2009,BoMi-Science-2012} and long-wavelength drivers are of high interest for molecular tomography~\cite{PaVi-PRL-2006,HaCa-JPhysB-2011,VoNe-NatPhys-2011}.
First steps to combine many-body approaches and strong-field physics in molecules have been made~\cite{AwVa-JPB-2005,ScSm-JCP126-2007,KlSa-JCP131-2009,SpPa-PRA-2009,ToIv-RPA-2012} but much more progress is needed before we are able to understand how we can control electronic motion in molecules as we starting to understand it in atomic systems~\cite{WiGo-Science-2011,OtKa-Science-2013}.

\acknowledgments
This work has been supported by the Deutsche Forschungsgemeinschaft (DFG) under grant No. SFB 925/A5.

\bibliographystyle{apsrev4-1-etal}
\bibliography{amo,books,notes}

\begin{thebibliography}{69}%
\makeatletter
\providecommand \@ifxundefined [1]{%
 \@ifx{#1\undefined}
}%
\providecommand \@ifnum [1]{%
 \ifnum #1\expandafter \@firstoftwo
 \else \expandafter \@secondoftwo
 \fi
}%
\providecommand \@ifx [1]{%
 \ifx #1\expandafter \@firstoftwo
 \else \expandafter \@secondoftwo
 \fi
}%
\providecommand \natexlab [1]{#1}%
\providecommand \enquote  [1]{``#1''}%
\providecommand \bibnamefont  [1]{#1}%
\providecommand \bibfnamefont [1]{#1}%
\providecommand \citenamefont [1]{#1}%
\providecommand \href@noop [0]{\@secondoftwo}%
\providecommand \href[0]{\begingroup \@sanitize@url \@href}%
\providecommand \@href[1]{\@@startlink{#1}\@@href}%
\providecommand \@@href[1]{\endgroup#1\@@endlink}%
\providecommand \@sanitize@url [0]{\catcode `\\12\catcode `\$12\catcode
  `\&12\catcode `\#12\catcode `\^12\catcode `\_12\catcode `\%12\relax}%
\providecommand \@@startlink[1]{}%
\providecommand \@@endlink[0]{}%
\providecommand \url  [0]{\begingroup\@sanitize@url \@url }%
\providecommand \@url [1]{\endgroup\@href {#1}{\urlprefix }}%
\providecommand \urlprefix  [0]{URL }%
\providecommand \Eprint [0]{\href }%
\providecommand \doibase [0]{http://dx.doi.org/}%
\providecommand \selectlanguage [0]{\@gobble}%
\providecommand \bibinfo  [0]{\@secondoftwo}%
\providecommand \bibfield  [0]{\@secondoftwo}%
\providecommand \translation [1]{[#1]}%
\providecommand \BibitemOpen [0]{}%
\providecommand \bibitemStop [0]{}%
\providecommand \bibitemNoStop [0]{.\EOS\space}%
\providecommand \EOS [0]{\spacefactor3000\relax}%
\providecommand \BibitemShut  [1]{\csname bibitem#1\endcsname}%
\let\auto@bib@innerbib\@empty
\bibitem [{\citenamefont {Krausz}\ and\ \citenamefont
  {Ivanov}(2009)}]{KrIv-RMP-2009}%
  \BibitemOpen
  \bibfield  {author} {\bibinfo {author} {\bibfnamefont {F.}~\bibnamefont
  {Krausz}}\ and\ \bibinfo {author} {\bibfnamefont {M.}~\bibnamefont
  {Ivanov}},\ }\href{\doibase 10.1103/RevModPhys.81.163} {\bibfield  {journal}
  {\bibinfo  {journal} {Rev. Mod. Phys.}\ }\textbf {\bibinfo {volume} {81}},\
  \bibinfo {pages} {163} (\bibinfo {year} {2009})}\BibitemShut {NoStop}%
\bibitem [{\citenamefont {Kohler}\ \emph {et~al.}(2012)\citenamefont {Kohler},
  \citenamefont {Pfeifer}, \citenamefont {Hatsagortsyan},\ and\ \citenamefont
  {Keitel}}]{KoPf-AAMO-2012}%
  \BibitemOpen
  \bibfield  {author} {\bibinfo {author} {\bibfnamefont {M.}~\bibnamefont
  {Kohler}}, \bibinfo {author} {\bibfnamefont {T.}~\bibnamefont {Pfeifer}},
  \bibinfo {author} {\bibfnamefont {K.}~\bibnamefont {Hatsagortsyan}}, \ and\
  \bibinfo {author} {\bibfnamefont {C.}~\bibnamefont {Keitel}},\ }\bibfield
  {booktitle} {\emph {\bibinfo {booktitle} {Advances in Atomic, Molecular, and
  Optical Physics}},\ }\href{\doibase 10.1016/B978-0-12-396482-3.00004-1}
  {\bibfield  {journal} {\bibinfo  {journal} {Adv. At., Mol., Opt. Phys.}\
  }\textbf {\bibinfo {volume} {61}},\ \bibinfo {pages} {159 } (\bibinfo {year}
  {2012})}\BibitemShut {NoStop}%
\bibitem [{\citenamefont {Pabst}(2013)}]{Pa-EPJST-2013}%
  \BibitemOpen
  \bibfield  {author} {\bibinfo {author} {\bibfnamefont {S.}~\bibnamefont
  {Pabst}},\ }\href{\doibase 10.1140/epjst/e2013-01819-x} {\bibfield  {journal}
  {\bibinfo  {journal} {Eur. Phys. J. Special Topics}\ }\textbf {\bibinfo
  {volume} {221}},\ \bibinfo {pages} {1} (\bibinfo {year} {2013})}\BibitemShut
  {NoStop}%
\bibitem [{\citenamefont {Agostini}\ and\ \citenamefont
  {DiMauro}(2004)}]{AgDi-RPP-2004}%
  \BibitemOpen
  \bibfield  {author} {\bibinfo {author} {\bibfnamefont {P.}~\bibnamefont
  {Agostini}}\ and\ \bibinfo {author} {\bibfnamefont {L.~F.}\ \bibnamefont
  {DiMauro}},\ }\href{http://stacks.iop.org/0034-4885/67/i=6/a=R01} {\bibfield
  {journal} {\bibinfo  {journal} {Reports on Progress in Physics}\ }\textbf
  {\bibinfo {volume} {67}},\ \bibinfo {pages} {813} (\bibinfo {year}
  {2004})}\BibitemShut {NoStop}%
\bibitem [{\citenamefont {Papadogiannis}\ \emph {et~al.}(1999)\citenamefont
  {Papadogiannis}, \citenamefont {Witzel}, \citenamefont {Kalpouzos},\ and\
  \citenamefont {Charalambidis}}]{PaCh-PRL-1999}%
  \BibitemOpen
  \bibfield  {author} {\bibinfo {author} {\bibfnamefont {N.~A.}\ \bibnamefont
  {Papadogiannis}}, \bibinfo {author} {\bibfnamefont {B.}~\bibnamefont
  {Witzel}}, \bibinfo {author} {\bibfnamefont {C.}~\bibnamefont {Kalpouzos}}, \
  and\ \bibinfo {author} {\bibfnamefont {D.}~\bibnamefont {Charalambidis}},\
  }\href{\doibase 10.1103/PhysRevLett.83.4289} {\bibfield  {journal} {\bibinfo
  {journal} {Phys. Rev. Lett.}\ }\textbf {\bibinfo {volume} {83}},\ \bibinfo
  {pages} {4289} (\bibinfo {year} {1999})}\BibitemShut {NoStop}%
\bibitem [{\citenamefont {Paul}\ \emph {et~al.}(2001)\citenamefont {Paul},
  \citenamefont {Toma}, \citenamefont {Breger}, \citenamefont {Mullot},
  \citenamefont {Aug\'e}, \citenamefont {Balcou}, \citenamefont {Muller},\ and\
  \citenamefont {Agostini}}]{PaTo-Science-2001}%
  \BibitemOpen
  \bibfield  {author} {\bibinfo {author} {\bibfnamefont {P.~M.}\ \bibnamefont
  {Paul}}, \bibinfo {author} {\bibfnamefont {E.~S.}\ \bibnamefont {Toma}},
  \bibinfo {author} {\bibfnamefont {P.}~\bibnamefont {Breger}},  \emph
  {et~al.},\ }\href{\doibase 10.1126/science.1059413} {\bibfield  {journal}
  {\bibinfo  {journal} {Science}\ }\textbf {\bibinfo {volume} {292}},\ \bibinfo
  {pages} {1689} (\bibinfo {year} {2001})}\BibitemShut {NoStop}%
\bibitem [{\citenamefont {Popmintchev}\ \emph {et~al.}(2012)\citenamefont
  {Popmintchev}, \citenamefont {Chen}, \citenamefont {Popmintchev},
  \citenamefont {Arpin}, \citenamefont {Brown}, \citenamefont {Ališauskas},
  \citenamefont {Andriukaitis}, \citenamefont {Bal\v{c}iunas}, \citenamefont
  {M\"ucke}, \citenamefont {Pugzlys}, \citenamefont {Baltui\v{s}ka},
  \citenamefont {Shim}, \citenamefont {Schrauth}, \citenamefont {Gaeta},
  \citenamefont {Hern\'andez-Garc\'ia}, \citenamefont {Plaja}, \citenamefont
  {Becker}, \citenamefont {Jaron-Becker}, \citenamefont {Murnane},\ and\
  \citenamefont {Kapteyn}}]{PoCh-Science-2012}%
  \BibitemOpen
  \bibfield  {author} {\bibinfo {author} {\bibfnamefont {T.}~\bibnamefont
  {Popmintchev}}, \bibinfo {author} {\bibfnamefont {M.-C.}\ \bibnamefont
  {Chen}}, \bibinfo {author} {\bibfnamefont {D.}~\bibnamefont {Popmintchev}},
  \emph {et~al.},\ }\href{\doibase 10.1126/science.1218497} {\bibfield
  {journal} {\bibinfo  {journal} {Science}\ }\textbf {\bibinfo {volume}
  {336}},\ \bibinfo {pages} {1287} (\bibinfo {year} {2012})}\BibitemShut
  {NoStop}%
\bibitem [{\citenamefont {Zhao}\ \emph {et~al.}(2012)\citenamefont {Zhao},
  \citenamefont {Zhang}, \citenamefont {Chini}, \citenamefont {Wu},
  \citenamefont {Wang},\ and\ \citenamefont {Chang}}]{ZhZh-OptLett-2012}%
  \BibitemOpen
  \bibfield  {author} {\bibinfo {author} {\bibfnamefont {K.}~\bibnamefont
  {Zhao}}, \bibinfo {author} {\bibfnamefont {Q.}~\bibnamefont {Zhang}},
  \bibinfo {author} {\bibfnamefont {M.}~\bibnamefont {Chini}}, \bibinfo
  {author} {\bibfnamefont {Y.}~\bibnamefont {Wu}}, \bibinfo {author}
  {\bibfnamefont {X.}~\bibnamefont {Wang}}, \ and\ \bibinfo {author}
  {\bibfnamefont {Z.}~\bibnamefont {Chang}},\
  }\href{http://ol.osa.org/abstract.cfm?URI=ol-37-18-3891} {\bibfield
  {journal} {\bibinfo  {journal} {Opt. Lett.}\ }\textbf {\bibinfo {volume}
  {37}},\ \bibinfo {pages} {3891} (\bibinfo {year} {2012})}\BibitemShut
  {NoStop}%
\bibitem [{\citenamefont {Goulielmakis}\ \emph {et~al.}(2010)\citenamefont
  {Goulielmakis}, \citenamefont {Loh}, \citenamefont {Wirth}, \citenamefont
  {Santra}, \citenamefont {Rohringer}, \citenamefont {Yakovlev}, \citenamefont
  {Zherebtsov}, \citenamefont {Pfeifer}, \citenamefont {Azzeer}, \citenamefont
  {Kling}, \citenamefont {Leone},\ and\ \citenamefont
  {Krausz}}]{GoKr-Nature-2010}%
  \BibitemOpen
  \bibfield  {author} {\bibinfo {author} {\bibfnamefont {E.}~\bibnamefont
  {Goulielmakis}}, \bibinfo {author} {\bibfnamefont {Z.}~\bibnamefont {Loh}},
  \bibinfo {author} {\bibfnamefont {A.}~\bibnamefont {Wirth}},  \emph
  {et~al.},\ }\href{\doibase 10.1038/nature09212} {\bibfield  {journal}
  {\bibinfo  {journal} {Nature}\ }\textbf {\bibinfo {volume} {466}},\ \bibinfo
  {pages} {739} (\bibinfo {year} {2010})}\BibitemShut {NoStop}%
\bibitem [{\citenamefont {Wirth}\ \emph {et~al.}(2011)\citenamefont {Wirth},
  \citenamefont {Hassan}, \citenamefont {Grguras}, \citenamefont {Gagnon},
  \citenamefont {Moulet}, \citenamefont {Luu}, \citenamefont {Pabst},
  \citenamefont {Santra}, \citenamefont {Alahmed}, \citenamefont {Azzeer},
  \citenamefont {Yakovlev}, \citenamefont {Pervak}, \citenamefont {Krausz},\
  and\ \citenamefont {Goulielmakis}}]{WiGo-Science-2011}%
  \BibitemOpen
  \bibfield  {author} {\bibinfo {author} {\bibfnamefont {A.}~\bibnamefont
  {Wirth}}, \bibinfo {author} {\bibfnamefont {M.~T.}\ \bibnamefont {Hassan}},
  \bibinfo {author} {\bibfnamefont {I.}~\bibnamefont {Grguras}},  \emph
  {et~al.},\ }\href{\doibase 10.1126/science.1210268} {\bibfield  {journal}
  {\bibinfo  {journal} {Science}\ }\textbf {\bibinfo {volume} {334}},\ \bibinfo
  {pages} {195} (\bibinfo {year} {2011})}\BibitemShut {NoStop}%
\bibitem [{\citenamefont {Ott}\ \emph {et~al.}(2013)\citenamefont {Ott},
  \citenamefont {Kaldun}, \citenamefont {Raith}, \citenamefont {Meyer},
  \citenamefont {Laux}, \citenamefont {Evers}, \citenamefont {Keitel},
  \citenamefont {Greene},\ and\ \citenamefont {Pfeifer}}]{OtKa-Science-2013}%
  \BibitemOpen
  \bibfield  {author} {\bibinfo {author} {\bibfnamefont {C.}~\bibnamefont
  {Ott}}, \bibinfo {author} {\bibfnamefont {A.}~\bibnamefont {Kaldun}},
  \bibinfo {author} {\bibfnamefont {P.}~\bibnamefont {Raith}},  \emph
  {et~al.},\ }\href{\doibase 10.1126/science.1234407} {\bibfield  {journal}
  {\bibinfo  {journal} {Science}\ }\textbf {\bibinfo {volume} {340}},\ \bibinfo
  {pages} {716} (\bibinfo {year} {2013})}\BibitemShut {NoStop}%
\bibitem [{\citenamefont {Drescher}\ \emph {et~al.}(2002)\citenamefont
  {Drescher}, \citenamefont {Hentschel}, \citenamefont {Kienberger},
  \citenamefont {Uiberacker}, \citenamefont {Yakovlev}, \citenamefont
  {Scrinzi}, \citenamefont {Westerwalbesloh}, \citenamefont {Kleineberg},
  \citenamefont {Heinzmann},\ and\ \citenamefont {Krausz}}]{DrKr-Nature-2002}%
  \BibitemOpen
  \bibfield  {author} {\bibinfo {author} {\bibfnamefont {M.}~\bibnamefont
  {Drescher}}, \bibinfo {author} {\bibfnamefont {M.}~\bibnamefont {Hentschel}},
  \bibinfo {author} {\bibfnamefont {R.}~\bibnamefont {Kienberger}},  \emph
  {et~al.},\ }\href{\doibase 10.1038/nature01143} {\bibfield  {journal}
  {\bibinfo  {journal} {Nature}\ }\textbf {\bibinfo {volume} {419}},\ \bibinfo
  {pages} {803} (\bibinfo {year} {2002})}\BibitemShut {NoStop}%
\bibitem [{\citenamefont {Itatani}\ \emph {et~al.}(2002)\citenamefont
  {Itatani}, \citenamefont {Qu\'er\'e}, \citenamefont {Yudin}, \citenamefont
  {Ivanov}, \citenamefont {Krausz},\ and\ \citenamefont
  {Corkum}}]{ItQu-PRL-2002}%
  \BibitemOpen
  \bibfield  {author} {\bibinfo {author} {\bibfnamefont {J.}~\bibnamefont
  {Itatani}}, \bibinfo {author} {\bibfnamefont {F.}~\bibnamefont {Qu\'er\'e}},
  \bibinfo {author} {\bibfnamefont {G.~L.}\ \bibnamefont {Yudin}}, \bibinfo
  {author} {\bibfnamefont {M.~Y.}\ \bibnamefont {Ivanov}}, \bibinfo {author}
  {\bibfnamefont {F.}~\bibnamefont {Krausz}}, \ and\ \bibinfo {author}
  {\bibfnamefont {P.~B.}\ \bibnamefont {Corkum}},\ }\href{\doibase
  10.1103/PhysRevLett.88.173903} {\bibfield  {journal} {\bibinfo  {journal}
  {Phys. Rev. Lett.}\ }\textbf {\bibinfo {volume} {88}},\ \bibinfo {pages}
  {173903} (\bibinfo {year} {2002})}\BibitemShut {NoStop}%
\bibitem [{\citenamefont {Schultze}\ \emph {et~al.}(2010)\citenamefont
  {Schultze}, \citenamefont {Fiess}, \citenamefont {Karpowicz}, \citenamefont
  {Gagnon}, \citenamefont {Korbman}, \citenamefont {Hofstetter}, \citenamefont
  {Neppl}, \citenamefont {Cavalieri}, \citenamefont {Komninos}, \citenamefont
  {Mercouris}, \citenamefont {Nicolaides}, \citenamefont {Pazourek},
  \citenamefont {Nagele}, \citenamefont {Feist}, \citenamefont {Burgdorfer},
  \citenamefont {Azzeer}, \citenamefont {Ernstorfer}, \citenamefont
  {Kienberger}, \citenamefont {Kleineberg}, \citenamefont {Goulielmakis},
  \citenamefont {Krausz},\ and\ \citenamefont {Yakovlev}}]{ScYa-Science-2010}%
  \BibitemOpen
  \bibfield  {author} {\bibinfo {author} {\bibfnamefont {M.}~\bibnamefont
  {Schultze}}, \bibinfo {author} {\bibfnamefont {M.}~\bibnamefont {Fiess}},
  \bibinfo {author} {\bibfnamefont {N.}~\bibnamefont {Karpowicz}},  \emph
  {et~al.},\ }\href{\doibase 10.1126/science.1189401} {\bibfield  {journal}
  {\bibinfo  {journal} {Science}\ }\textbf {\bibinfo {volume} {328}},\ \bibinfo
  {pages} {1658} (\bibinfo {year} {2010})}\BibitemShut {NoStop}%
\bibitem [{\citenamefont {K\"{o}hler}\ \emph {et~al.}(2011)\citenamefont
  {K\"{o}hler}, \citenamefont {Wollenhaupt}, \citenamefont {Bayer},
  \citenamefont {Sarpe},\ and\ \citenamefont {Baumert}}]{KoWo-OptExp-2011}%
  \BibitemOpen
  \bibfield  {author} {\bibinfo {author} {\bibfnamefont {J.}~\bibnamefont
  {K\"{o}hler}}, \bibinfo {author} {\bibfnamefont {M.}~\bibnamefont
  {Wollenhaupt}}, \bibinfo {author} {\bibfnamefont {T.}~\bibnamefont {Bayer}},
  \bibinfo {author} {\bibfnamefont {C.}~\bibnamefont {Sarpe}}, \ and\ \bibinfo
  {author} {\bibfnamefont {T.}~\bibnamefont {Baumert}},\ }\href{\doibase
  10.1364/OE.19.011638} {\bibfield  {journal} {\bibinfo  {journal} {Opt.
  Express}\ }\textbf {\bibinfo {volume} {19}},\ \bibinfo {pages} {11638}
  (\bibinfo {year} {2011})}\BibitemShut {NoStop}%
\bibitem [{\citenamefont {Itatani}\ \emph {et~al.}(2004)\citenamefont
  {Itatani}, \citenamefont {Levesque}, \citenamefont {Zeidler}, \citenamefont
  {Niikura}, \citenamefont {Pepin}, \citenamefont {Kieffer}, \citenamefont
  {Corkum},\ and\ \citenamefont {Villeneuve}}]{ItLe-Nature432-2004}%
  \BibitemOpen
  \bibfield  {author} {\bibinfo {author} {\bibfnamefont {J.}~\bibnamefont
  {Itatani}}, \bibinfo {author} {\bibfnamefont {J.}~\bibnamefont {Levesque}},
  \bibinfo {author} {\bibfnamefont {D.}~\bibnamefont {Zeidler}},  \emph
  {et~al.},\ }\href{\doibase 10.1038/nature03183} {\bibfield  {journal}
  {\bibinfo  {journal} {Nature}\ }\textbf {\bibinfo {volume} {432}},\ \bibinfo
  {pages} {867} (\bibinfo {year} {2004})}\BibitemShut {NoStop}%
\bibitem [{\citenamefont {McFarland}\ \emph {et~al.}(2008)\citenamefont
  {McFarland}, \citenamefont {Farrell}, \citenamefont {Bucksbaum},\ and\
  \citenamefont {G\"uhr}}]{McGu-Science-2008}%
  \BibitemOpen
  \bibfield  {author} {\bibinfo {author} {\bibfnamefont {B.~K.}\ \bibnamefont
  {McFarland}}, \bibinfo {author} {\bibfnamefont {J.~P.}\ \bibnamefont
  {Farrell}}, \bibinfo {author} {\bibfnamefont {P.~H.}\ \bibnamefont
  {Bucksbaum}}, \ and\ \bibinfo {author} {\bibfnamefont {M.}~\bibnamefont
  {G\"uhr}},\ }\href{\doibase 10.1126/science.1162780} {\bibfield  {journal}
  {\bibinfo  {journal} {Science}\ }\textbf {\bibinfo {volume} {322}},\ \bibinfo
  {pages} {1232} (\bibinfo {year} {2008})}\BibitemShut {NoStop}%
\bibitem [{\citenamefont {Smirnova}\ \emph {et~al.}(2009)\citenamefont
  {Smirnova}, \citenamefont {Mairesse}, \citenamefont {Patchkovskii},
  \citenamefont {Dudovich}, \citenamefont {Villeneuve}, \citenamefont
  {Corkum},\ and\ \citenamefont {Ivanov}}]{SmIv-Nature-2009}%
  \BibitemOpen
  \bibfield  {author} {\bibinfo {author} {\bibfnamefont {O.}~\bibnamefont
  {Smirnova}}, \bibinfo {author} {\bibfnamefont {Y.}~\bibnamefont {Mairesse}},
  \bibinfo {author} {\bibfnamefont {S.}~\bibnamefont {Patchkovskii}}, \bibinfo
  {author} {\bibfnamefont {N.}~\bibnamefont {Dudovich}}, \bibinfo {author}
  {\bibfnamefont {D.}~\bibnamefont {Villeneuve}}, \bibinfo {author}
  {\bibfnamefont {P.}~\bibnamefont {Corkum}}, \ and\ \bibinfo {author}
  {\bibfnamefont {M.~Y.}\ \bibnamefont {Ivanov}},\ }\href{\doibase
  10.1038/nature08253} {\bibfield  {journal} {\bibinfo  {journal} {Nature}\
  }\textbf {\bibinfo {volume} {460}},\ \bibinfo {pages} {972} (\bibinfo {year}
  {2009})}\BibitemShut {NoStop}%
\bibitem [{\citenamefont {Haessler}\ \emph {et~al.}(2011)\citenamefont
  {Haessler}, \citenamefont {Caillat},\ and\ \citenamefont
  {Sali\'eres}}]{HaCa-JPhysB-2011}%
  \BibitemOpen
  \bibfield  {author} {\bibinfo {author} {\bibfnamefont {S.}~\bibnamefont
  {Haessler}}, \bibinfo {author} {\bibfnamefont {J.}~\bibnamefont {Caillat}}, \
  and\ \bibinfo {author} {\bibfnamefont {P.}~\bibnamefont {Sali\'eres}},\
  }\href{http://stacks.iop.org/0953-4075/44/i=20/a=203001} {\bibfield
  {journal} {\bibinfo  {journal} {J. Phys. B}\ }\textbf {\bibinfo {volume}
  {44}},\ \bibinfo {pages} {203001} (\bibinfo {year} {2011})}\BibitemShut
  {NoStop}%
\bibitem [{\citenamefont {Shafir}\ \emph {et~al.}(2012)\citenamefont {Shafir},
  \citenamefont {Soifer}, \citenamefont {Bruner}, \citenamefont {Dagan},
  \citenamefont {Mairesse}, \citenamefont {Patchkovskii}, \citenamefont
  {Ivanov}, \citenamefont {Smirnova},\ and\ \citenamefont
  {Dudovich}}]{ShSo-Nature-2012}%
  \BibitemOpen
  \bibfield  {author} {\bibinfo {author} {\bibfnamefont {D.}~\bibnamefont
  {Shafir}}, \bibinfo {author} {\bibfnamefont {H.}~\bibnamefont {Soifer}},
  \bibinfo {author} {\bibfnamefont {B.~D.}\ \bibnamefont {Bruner}},  \emph
  {et~al.},\ }\href{http://dx.doi.org/10.1038/nature11025} {\bibfield
  {journal} {\bibinfo  {journal} {Nature}\ }\textbf {\bibinfo {volume} {485}},\
  \bibinfo {pages} {343} (\bibinfo {year} {2012})}\BibitemShut {NoStop}%
\bibitem [{\citenamefont {Shivaram}\ \emph {et~al.}(2012)\citenamefont
  {Shivaram}, \citenamefont {Timmers}, \citenamefont {Tong},\ and\
  \citenamefont {Sandhu}}]{ShTi-PRL-2012}%
  \BibitemOpen
  \bibfield  {author} {\bibinfo {author} {\bibfnamefont {N.}~\bibnamefont
  {Shivaram}}, \bibinfo {author} {\bibfnamefont {H.}~\bibnamefont {Timmers}},
  \bibinfo {author} {\bibfnamefont {X.-M.}\ \bibnamefont {Tong}}, \ and\
  \bibinfo {author} {\bibfnamefont {A.}~\bibnamefont {Sandhu}},\
  }\href{\doibase 10.1103/PhysRevLett.108.193002} {\bibfield  {journal}
  {\bibinfo  {journal} {Phys. Rev. Lett.}\ }\textbf {\bibinfo {volume} {108}},\
  \bibinfo {pages} {193002} (\bibinfo {year} {2012})}\BibitemShut {NoStop}%
\bibitem [{\citenamefont {W\"orner}\ \emph {et~al.}(2009)\citenamefont
  {W\"orner}, \citenamefont {Niikura}, \citenamefont {Bertrand}, \citenamefont
  {Corkum},\ and\ \citenamefont {Villeneuve}}]{WoVi-PRL-2009}%
  \BibitemOpen
  \bibfield  {author} {\bibinfo {author} {\bibfnamefont {H.~J.}\ \bibnamefont
  {W\"orner}}, \bibinfo {author} {\bibfnamefont {H.}~\bibnamefont {Niikura}},
  \bibinfo {author} {\bibfnamefont {J.~B.}\ \bibnamefont {Bertrand}}, \bibinfo
  {author} {\bibfnamefont {P.~B.}\ \bibnamefont {Corkum}}, \ and\ \bibinfo
  {author} {\bibfnamefont {D.~M.}\ \bibnamefont {Villeneuve}},\ }\href{\doibase
  10.1103/PhysRevLett.102.103901} {\bibfield  {journal} {\bibinfo  {journal}
  {Phys. Rev. Lett.}\ }\textbf {\bibinfo {volume} {102}},\ \bibinfo {pages}
  {103901} (\bibinfo {year} {2009})}\BibitemShut {NoStop}%
\bibitem [{\citenamefont {Higuet}\ \emph {et~al.}(2011)\citenamefont {Higuet},
  \citenamefont {Ruf}, \citenamefont {Thir\'e}, \citenamefont {Cireasa},
  \citenamefont {Constant}, \citenamefont {Cormier}, \citenamefont {Descamps},
  \citenamefont {M\'evel}, \citenamefont {Petit}, \citenamefont {Pons},
  \citenamefont {Mairesse},\ and\ \citenamefont {Fabre}}]{HiFa-PRA-2011}%
  \BibitemOpen
  \bibfield  {author} {\bibinfo {author} {\bibfnamefont {J.}~\bibnamefont
  {Higuet}}, \bibinfo {author} {\bibfnamefont {H.}~\bibnamefont {Ruf}},
  \bibinfo {author} {\bibfnamefont {N.}~\bibnamefont {Thir\'e}},  \emph
  {et~al.},\ }\href{\doibase 10.1103/PhysRevA.83.053401} {\bibfield  {journal}
  {\bibinfo  {journal} {Phys. Rev. A}\ }\textbf {\bibinfo {volume} {83}},\
  \bibinfo {pages} {053401} (\bibinfo {year} {2011})}\BibitemShut {NoStop}%
\bibitem [{\citenamefont {Farrell}\ \emph {et~al.}(2011)\citenamefont
  {Farrell}, \citenamefont {Spector}, \citenamefont {McFarland}, \citenamefont
  {Bucksbaum}, \citenamefont {G\"uhr}, \citenamefont {Gaarde},\ and\
  \citenamefont {Schafer}}]{FaSc-PRA-2011}%
  \BibitemOpen
  \bibfield  {author} {\bibinfo {author} {\bibfnamefont {J.~P.}\ \bibnamefont
  {Farrell}}, \bibinfo {author} {\bibfnamefont {L.~S.}\ \bibnamefont
  {Spector}}, \bibinfo {author} {\bibfnamefont {B.~K.}\ \bibnamefont
  {McFarland}}, \bibinfo {author} {\bibfnamefont {P.~H.}\ \bibnamefont
  {Bucksbaum}}, \bibinfo {author} {\bibfnamefont {M.}~\bibnamefont {G\"uhr}},
  \bibinfo {author} {\bibfnamefont {M.~B.}\ \bibnamefont {Gaarde}}, \ and\
  \bibinfo {author} {\bibfnamefont {K.~J.}\ \bibnamefont {Schafer}},\
  }\href{\doibase 10.1103/PhysRevA.83.023420} {\bibfield  {journal} {\bibinfo
  {journal} {Phys. Rev. A}\ }\textbf {\bibinfo {volume} {83}},\ \bibinfo
  {pages} {023420} (\bibinfo {year} {2011})}\BibitemShut {NoStop}%
\bibitem [{\citenamefont {Pabst}\ \emph
  {et~al.}(2012{\natexlab{a}})\citenamefont {Pabst}, \citenamefont {Greenman},
  \citenamefont {Mazziotti},\ and\ \citenamefont {Santra}}]{PaGr-PRA-2012}%
  \BibitemOpen
  \bibfield  {author} {\bibinfo {author} {\bibfnamefont {S.}~\bibnamefont
  {Pabst}}, \bibinfo {author} {\bibfnamefont {L.}~\bibnamefont {Greenman}},
  \bibinfo {author} {\bibfnamefont {D.~A.}\ \bibnamefont {Mazziotti}}, \ and\
  \bibinfo {author} {\bibfnamefont {R.}~\bibnamefont {Santra}},\
  }\href{\doibase 10.1103/PhysRevA.85.023411} {\bibfield  {journal} {\bibinfo
  {journal} {Phys. Rev. A}\ }\textbf {\bibinfo {volume} {85}},\ \bibinfo
  {pages} {023411} (\bibinfo {year} {2012}{\natexlab{a}})}\BibitemShut
  {NoStop}%
\bibitem [{\citenamefont {Shiner}\ \emph {et~al.}(2011)\citenamefont {Shiner},
  \citenamefont {Schmidt}, \citenamefont {{Trallero-Herrero}}, \citenamefont
  {W\"orner}, \citenamefont {Patchkovskii}, \citenamefont {Corkum},
  \citenamefont {Kieffer}, \citenamefont {Legare},\ and\ \citenamefont
  {Villeneuve}}]{ShVi-NatPhys-2011}%
  \BibitemOpen
  \bibfield  {author} {\bibinfo {author} {\bibfnamefont {A.~D.}\ \bibnamefont
  {Shiner}}, \bibinfo {author} {\bibfnamefont {B.~E.}\ \bibnamefont {Schmidt}},
  \bibinfo {author} {\bibfnamefont {C.}~\bibnamefont {{Trallero-Herrero}}},
  \emph {et~al.},\ }\href{\doibase 10.1038/nphys1940} {\bibfield  {journal}
  {\bibinfo  {journal} {Nat. Phys.}\ }\textbf {\bibinfo {volume} {7}},\
  \bibinfo {pages} {464} (\bibinfo {year} {2011})}\BibitemShut {NoStop}%
\bibitem [{\citenamefont {Zhang}\ and\ \citenamefont
  {Guo}(2013)}]{ZhGu-PRL-2013}%
  \BibitemOpen
  \bibfield  {author} {\bibinfo {author} {\bibfnamefont {J.}~\bibnamefont
  {Zhang}}\ and\ \bibinfo {author} {\bibfnamefont {D.-S.}\ \bibnamefont
  {Guo}},\ }\href{\doibase 10.1103/PhysRevLett.110.063002} {\bibfield
  {journal} {\bibinfo  {journal} {Phys. Rev. Lett.}\ }\textbf {\bibinfo
  {volume} {110}},\ \bibinfo {pages} {063002} (\bibinfo {year}
  {2013})}\BibitemShut {NoStop}%
\bibitem [{\citenamefont {Vozzi}\ \emph
  {et~al.}(2011{\natexlab{a}})\citenamefont {Vozzi}, \citenamefont {Negro},
  \citenamefont {Calegari}, \citenamefont {Stagira}, \citenamefont {Kov\'acs},\
  and\ \citenamefont {Tosa}}]{VoNe-NJP-2011}%
  \BibitemOpen
  \bibfield  {author} {\bibinfo {author} {\bibfnamefont {C.}~\bibnamefont
  {Vozzi}}, \bibinfo {author} {\bibfnamefont {M.}~\bibnamefont {Negro}},
  \bibinfo {author} {\bibfnamefont {F.}~\bibnamefont {Calegari}}, \bibinfo
  {author} {\bibfnamefont {S.}~\bibnamefont {Stagira}}, \bibinfo {author}
  {\bibfnamefont {K.}~\bibnamefont {Kov\'acs}}, \ and\ \bibinfo {author}
  {\bibfnamefont {V.}~\bibnamefont {Tosa}},\
  }\href{http://stacks.iop.org/1367-2630/13/i=7/a=073003} {\bibfield  {journal}
  {\bibinfo  {journal} {New J. Phys.}\ }\textbf {\bibinfo {volume} {13}},\
  \bibinfo {pages} {073003} (\bibinfo {year} {2011}{\natexlab{a}})}\BibitemShut
  {NoStop}%
\bibitem [{\citenamefont {Sukharev}\ \emph {et~al.}(2002)\citenamefont
  {Sukharev}, \citenamefont {Charron},\ and\ \citenamefont
  {Suzor-Weiner}}]{SuCh-PRA-2002}%
  \BibitemOpen
  \bibfield  {author} {\bibinfo {author} {\bibfnamefont {M.}~\bibnamefont
  {Sukharev}}, \bibinfo {author} {\bibfnamefont {E.}~\bibnamefont {Charron}}, \
  and\ \bibinfo {author} {\bibfnamefont {A.}~\bibnamefont {Suzor-Weiner}},\
  }\href{\doibase 10.1103/PhysRevA.66.053407} {\bibfield  {journal} {\bibinfo
  {journal} {Phys. Rev. A}\ }\textbf {\bibinfo {volume} {66}},\ \bibinfo
  {pages} {053407} (\bibinfo {year} {2002})}\BibitemShut {NoStop}%
\bibitem [{\citenamefont {Scrinzi}(2012)}]{Sc-NJP-2012}%
  \BibitemOpen
  \bibfield  {author} {\bibinfo {author} {\bibfnamefont {A.}~\bibnamefont
  {Scrinzi}},\ }\href{http://stacks.iop.org/1367-2630/14/i=8/a=085008}
  {\bibfield  {journal} {\bibinfo  {journal} {New Journal of Physics}\ }\textbf
  {\bibinfo {volume} {14}},\ \bibinfo {pages} {085008} (\bibinfo {year}
  {2012})}\BibitemShut {NoStop}%
\bibitem [{\citenamefont {Miyagi}\ and\ \citenamefont
  {Madsen}(2013)}]{MiMa-PRA-2013}%
  \BibitemOpen
  \bibfield  {author} {\bibinfo {author} {\bibfnamefont {H.}~\bibnamefont
  {Miyagi}}\ and\ \bibinfo {author} {\bibfnamefont {L.~B.}\ \bibnamefont
  {Madsen}},\ }\href{\doibase 10.1103/PhysRevA.87.062511} {\bibfield  {journal}
  {\bibinfo  {journal} {Phys. Rev. A}\ }\textbf {\bibinfo {volume} {87}},\
  \bibinfo {pages} {062511} (\bibinfo {year} {2013})}\BibitemShut {NoStop}%
\bibitem [{\citenamefont {Scrinzi}\ and\ \citenamefont
  {Piraux}(1998)}]{ScPi-PRA-1998}%
  \BibitemOpen
  \bibfield  {author} {\bibinfo {author} {\bibfnamefont {A.}~\bibnamefont
  {Scrinzi}}\ and\ \bibinfo {author} {\bibfnamefont {B.}~\bibnamefont
  {Piraux}},\ }\href{\doibase 10.1103/PhysRevA.58.1310} {\bibfield  {journal}
  {\bibinfo  {journal} {Phys. Rev. A}\ }\textbf {\bibinfo {volume} {58}},\
  \bibinfo {pages} {1310} (\bibinfo {year} {1998})}\BibitemShut {NoStop}%
\bibitem [{\citenamefont {Prager}\ \emph {et~al.}(2001)\citenamefont {Prager},
  \citenamefont {Hu},\ and\ \citenamefont {Keitel}}]{PrHu-PRA-2001}%
  \BibitemOpen
  \bibfield  {author} {\bibinfo {author} {\bibfnamefont {J.}~\bibnamefont
  {Prager}}, \bibinfo {author} {\bibfnamefont {S.~X.}\ \bibnamefont {Hu}}, \
  and\ \bibinfo {author} {\bibfnamefont {C.~H.}\ \bibnamefont {Keitel}},\
  }\href{\doibase 10.1103/PhysRevA.64.045402} {\bibfield  {journal} {\bibinfo
  {journal} {Phys. Rev. A}\ }\textbf {\bibinfo {volume} {64}},\ \bibinfo
  {pages} {045402} (\bibinfo {year} {2001})}\BibitemShut {NoStop}%
\bibitem [{\citenamefont {Parker}\ \emph {et~al.}(2007)\citenamefont {Parker},
  \citenamefont {Meharg}, \citenamefont {McKenna},\ and\ \citenamefont
  {Taylor}}]{PaMe-JPB-2007}%
  \BibitemOpen
  \bibfield  {author} {\bibinfo {author} {\bibfnamefont {J.~S.}\ \bibnamefont
  {Parker}}, \bibinfo {author} {\bibfnamefont {K.~J.}\ \bibnamefont {Meharg}},
  \bibinfo {author} {\bibfnamefont {G.~A.}\ \bibnamefont {McKenna}}, \ and\
  \bibinfo {author} {\bibfnamefont {K.~T.}\ \bibnamefont {Taylor}},\
  }\href{http://stacks.iop.org/0953-4075/40/i=10/a=008} {\bibfield  {journal}
  {\bibinfo  {journal} {J. Phys. B: At., Mol. Opt. Phys.}\ }\textbf {\bibinfo
  {volume} {40}},\ \bibinfo {pages} {1729} (\bibinfo {year}
  {2007})}\BibitemShut {NoStop}%
\bibitem [{\citenamefont {Ngoko~Djiokap}\ and\ \citenamefont
  {Starace}(2011)}]{DjSt-PRA-2011}%
  \BibitemOpen
  \bibfield  {author} {\bibinfo {author} {\bibfnamefont {J.~M.}\ \bibnamefont
  {Ngoko~Djiokap}}\ and\ \bibinfo {author} {\bibfnamefont {A.~F.}\ \bibnamefont
  {Starace}},\ }\href{\doibase 10.1103/PhysRevA.84.013404} {\bibfield
  {journal} {\bibinfo  {journal} {Phys. Rev. A}\ }\textbf {\bibinfo {volume}
  {84}},\ \bibinfo {pages} {013404} (\bibinfo {year} {2011})}\BibitemShut
  {NoStop}%
\bibitem [{\citenamefont {Tarana}\ and\ \citenamefont
  {Greene}(2012)}]{TaGr-PRA-2012}%
  \BibitemOpen
  \bibfield  {author} {\bibinfo {author} {\bibfnamefont {M.}~\bibnamefont
  {Tarana}}\ and\ \bibinfo {author} {\bibfnamefont {C.~H.}\ \bibnamefont
  {Greene}},\ }\href{\doibase 10.1103/PhysRevA.85.013411} {\bibfield  {journal}
  {\bibinfo  {journal} {Phys. Rev. A}\ }\textbf {\bibinfo {volume} {85}},\
  \bibinfo {pages} {013411} (\bibinfo {year} {2012})}\BibitemShut {NoStop}%
\bibitem [{\citenamefont {Hochstuhl}\ and\ \citenamefont
  {Bonitz}(2012)}]{HoBo-PRA-2012}%
  \BibitemOpen
  \bibfield  {author} {\bibinfo {author} {\bibfnamefont {D.}~\bibnamefont
  {Hochstuhl}}\ and\ \bibinfo {author} {\bibfnamefont {M.}~\bibnamefont
  {Bonitz}},\ }\href{\doibase 10.1103/PhysRevA.86.053424} {\bibfield  {journal}
  {\bibinfo  {journal} {Phys. Rev. A}\ }\textbf {\bibinfo {volume} {86}},\
  \bibinfo {pages} {053424} (\bibinfo {year} {2012})}\BibitemShut {NoStop}%
\bibitem [{\citenamefont {Sanz-Vicario}\ \emph {et~al.}(2007)\citenamefont
  {Sanz-Vicario}, \citenamefont {Palacios}, \citenamefont {Cardona},
  \citenamefont {Bachau},\ and\ \citenamefont {Martin}}]{SaPa-JESRP-2007}%
  \BibitemOpen
  \bibfield  {author} {\bibinfo {author} {\bibfnamefont {J.}~\bibnamefont
  {Sanz-Vicario}}, \bibinfo {author} {\bibfnamefont {A.}~\bibnamefont
  {Palacios}}, \bibinfo {author} {\bibfnamefont {J.}~\bibnamefont {Cardona}},
  \bibinfo {author} {\bibfnamefont {H.}~\bibnamefont {Bachau}}, \ and\ \bibinfo
  {author} {\bibfnamefont {F.}~\bibnamefont {Martin}},\ }\href{\doibase
  http://dx.doi.org/10.1016/j.elspec.2007.02.011} {\bibfield  {journal}
  {\bibinfo  {journal} {Journal of Electron Spectroscopy and Related
  Phenomena}\ }\textbf {\bibinfo {volume} {161}},\ \bibinfo {pages} {182 }
  (\bibinfo {year} {2007})}\BibitemShut {NoStop}%
\bibitem [{\citenamefont {Brown}\ \emph
  {et~al.}(2012{\natexlab{a}})\citenamefont {Brown}, \citenamefont {Robinson},\
  and\ \citenamefont {van~der Hart}}]{BrRo-PRA-2012}%
  \BibitemOpen
  \bibfield  {author} {\bibinfo {author} {\bibfnamefont {A.~C.}\ \bibnamefont
  {Brown}}, \bibinfo {author} {\bibfnamefont {D.~J.}\ \bibnamefont {Robinson}},
  \ and\ \bibinfo {author} {\bibfnamefont {H.~W.}\ \bibnamefont {van~der
  Hart}},\ }\href{\doibase 10.1103/PhysRevA.86.053420} {\bibfield  {journal}
  {\bibinfo  {journal} {Phys. Rev. A}\ }\textbf {\bibinfo {volume} {86}},\
  \bibinfo {pages} {053420} (\bibinfo {year} {2012}{\natexlab{a}})}\BibitemShut
  {NoStop}%
\bibitem [{\citenamefont {Brown}\ \emph
  {et~al.}(2012{\natexlab{b}})\citenamefont {Brown}, \citenamefont
  {Hutchinson}, \citenamefont {Lysaght},\ and\ \citenamefont {van~der
  Hart}}]{BrHu-PRL-2012}%
  \BibitemOpen
  \bibfield  {author} {\bibinfo {author} {\bibfnamefont {A.~C.}\ \bibnamefont
  {Brown}}, \bibinfo {author} {\bibfnamefont {S.}~\bibnamefont {Hutchinson}},
  \bibinfo {author} {\bibfnamefont {M.~A.}\ \bibnamefont {Lysaght}}, \ and\
  \bibinfo {author} {\bibfnamefont {H.~W.}\ \bibnamefont {van~der Hart}},\
  }\href{\doibase 10.1103/PhysRevLett.108.063006} {\bibfield  {journal}
  {\bibinfo  {journal} {Phys. Rev. Lett.}\ }\textbf {\bibinfo {volume} {108}},\
  \bibinfo {pages} {063006} (\bibinfo {year} {2012}{\natexlab{b}})}\BibitemShut
  {NoStop}%
\bibitem [{n.h()}]{n.hugo}%
  \BibitemOpen
  \href@noop {} {}\bibinfo {note} {Personal communication with Prof. Hugo van
  der Hart.}\BibitemShut {Stop}%
\bibitem [{n.x()}]{n.xcid}%
  \BibitemOpen
  \href@noop {} {}\bibinfo {note} {S. Pabst, L. Greenman, and R. Santra --
  \textsc{xcid} program package for multichannel ionization dynamics, DESY,
  Hamburg, Germany, 2013, Rev. 792, with contributions from P. J. Ho., A.
  Sytcheva, and A. Karamatskou.}\BibitemShut {Stop}%
\bibitem [{\citenamefont {Greenman}\ \emph {et~al.}(2010)\citenamefont
  {Greenman}, \citenamefont {Ho}, \citenamefont {Pabst}, \citenamefont
  {Kamarchik}, \citenamefont {Mazziotti},\ and\ \citenamefont
  {Santra}}]{GrSa-PRA-2010}%
  \BibitemOpen
  \bibfield  {author} {\bibinfo {author} {\bibfnamefont {L.}~\bibnamefont
  {Greenman}}, \bibinfo {author} {\bibfnamefont {P.~J.}\ \bibnamefont {Ho}},
  \bibinfo {author} {\bibfnamefont {S.}~\bibnamefont {Pabst}}, \bibinfo
  {author} {\bibfnamefont {E.}~\bibnamefont {Kamarchik}}, \bibinfo {author}
  {\bibfnamefont {D.~A.}\ \bibnamefont {Mazziotti}}, \ and\ \bibinfo {author}
  {\bibfnamefont {R.}~\bibnamefont {Santra}},\ }\href{\doibase
  10.1103/PhysRevA.82.023406} {\bibfield  {journal} {\bibinfo  {journal} {Phys.
  Rev. A}\ }\textbf {\bibinfo {volume} {82}},\ \bibinfo {pages} {023406}
  (\bibinfo {year} {2010})}\BibitemShut {NoStop}%
\bibitem [{\citenamefont {Pabst}\ \emph {et~al.}(2011)\citenamefont {Pabst},
  \citenamefont {Greenman}, \citenamefont {Ho}, \citenamefont {Mazziotti},\
  and\ \citenamefont {Santra}}]{PaSa-PRL-2011}%
  \BibitemOpen
  \bibfield  {author} {\bibinfo {author} {\bibfnamefont {S.}~\bibnamefont
  {Pabst}}, \bibinfo {author} {\bibfnamefont {L.}~\bibnamefont {Greenman}},
  \bibinfo {author} {\bibfnamefont {P.~J.}\ \bibnamefont {Ho}}, \bibinfo
  {author} {\bibfnamefont {D.~A.}\ \bibnamefont {Mazziotti}}, \ and\ \bibinfo
  {author} {\bibfnamefont {R.}~\bibnamefont {Santra}},\ }\href{\doibase
  10.1103/PhysRevLett.106.053003} {\bibfield  {journal} {\bibinfo  {journal}
  {Phys. Rev. Lett.}\ }\textbf {\bibinfo {volume} {106}},\ \bibinfo {pages}
  {053003} (\bibinfo {year} {2011})}\BibitemShut {NoStop}%
\bibitem [{\citenamefont {Sytcheva}\ \emph {et~al.}(2012)\citenamefont
  {Sytcheva}, \citenamefont {Pabst}, \citenamefont {Son},\ and\ \citenamefont
  {Santra}}]{SyPa-PRA-2012}%
  \BibitemOpen
  \bibfield  {author} {\bibinfo {author} {\bibfnamefont {A.}~\bibnamefont
  {Sytcheva}}, \bibinfo {author} {\bibfnamefont {S.}~\bibnamefont {Pabst}},
  \bibinfo {author} {\bibfnamefont {S.-K.}\ \bibnamefont {Son}}, \ and\
  \bibinfo {author} {\bibfnamefont {R.}~\bibnamefont {Santra}},\
  }\href{\doibase 10.1103/PhysRevA.85.023414} {\bibfield  {journal} {\bibinfo
  {journal} {Phys. Rev. A}\ }\textbf {\bibinfo {volume} {85}},\ \bibinfo
  {pages} {023414} (\bibinfo {year} {2012})}\BibitemShut {NoStop}%
\bibitem [{\citenamefont {Pabst}\ \emph
  {et~al.}(2012{\natexlab{b}})\citenamefont {Pabst}, \citenamefont {Sytcheva},
  \citenamefont {Moulet}, \citenamefont {Wirth}, \citenamefont {Goulielmakis},\
  and\ \citenamefont {Santra}}]{PaSy-PRA-2012}%
  \BibitemOpen
  \bibfield  {author} {\bibinfo {author} {\bibfnamefont {S.}~\bibnamefont
  {Pabst}}, \bibinfo {author} {\bibfnamefont {A.}~\bibnamefont {Sytcheva}},
  \bibinfo {author} {\bibfnamefont {A.}~\bibnamefont {Moulet}}, \bibinfo
  {author} {\bibfnamefont {A.}~\bibnamefont {Wirth}}, \bibinfo {author}
  {\bibfnamefont {E.}~\bibnamefont {Goulielmakis}}, \ and\ \bibinfo {author}
  {\bibfnamefont {R.}~\bibnamefont {Santra}},\ }\href{\doibase
  10.1103/PhysRevA.86.063411} {\bibfield  {journal} {\bibinfo  {journal} {Phys.
  Rev. A}\ }\textbf {\bibinfo {volume} {86}},\ \bibinfo {pages} {063411}
  (\bibinfo {year} {2012}{\natexlab{b}})}\BibitemShut {NoStop}%
\bibitem [{\citenamefont {Karamatskou}\ \emph {et~al.}(2013)\citenamefont
  {Karamatskou}, \citenamefont {Pabst},\ and\ \citenamefont
  {Santra}}]{KaPa-PRA-2013}%
  \BibitemOpen
  \bibfield  {author} {\bibinfo {author} {\bibfnamefont {A.}~\bibnamefont
  {Karamatskou}}, \bibinfo {author} {\bibfnamefont {S.}~\bibnamefont {Pabst}},
  \ and\ \bibinfo {author} {\bibfnamefont {R.}~\bibnamefont {Santra}},\
  }\href{\doibase 10.1103/PhysRevA.87.043422} {\bibfield  {journal} {\bibinfo
  {journal} {Phys. Rev. A}\ }\textbf {\bibinfo {volume} {87}},\ \bibinfo
  {pages} {043422} (\bibinfo {year} {2013})}\BibitemShut {NoStop}%
\bibitem [{\citenamefont {Rohringer}\ \emph {et~al.}(2006)\citenamefont
  {Rohringer}, \citenamefont {Gordon},\ and\ \citenamefont
  {Santra}}]{RoSa-PRA-2006}%
  \BibitemOpen
  \bibfield  {author} {\bibinfo {author} {\bibfnamefont {N.}~\bibnamefont
  {Rohringer}}, \bibinfo {author} {\bibfnamefont {A.}~\bibnamefont {Gordon}}, \
  and\ \bibinfo {author} {\bibfnamefont {R.}~\bibnamefont {Santra}},\
  }\href{\doibase 10.1103/PhysRevA.74.043420} {\bibfield  {journal} {\bibinfo
  {journal} {Phys. Rev. A}\ }\textbf {\bibinfo {volume} {74}},\ \bibinfo
  {pages} {043420} (\bibinfo {year} {2006})}\BibitemShut {NoStop}%
\bibitem [{\citenamefont {Szabo}\ and\ \citenamefont
  {Ostlund}(1996)}]{SzOs-book}%
  \BibitemOpen
  \bibfield  {author} {\bibinfo {author} {\bibfnamefont {A.}~\bibnamefont
  {Szabo}}\ and\ \bibinfo {author} {\bibfnamefont {N.~S.}\ \bibnamefont
  {Ostlund}},\ }\href@noop {} {\emph {\bibinfo {title} {Modern Quantum
  Chemistry}}}\ (\bibinfo  {publisher} {Dover Publication Inc., Mineola, NY},\
  \bibinfo {year} {1996})\BibitemShut {NoStop}%
\bibitem [{\citenamefont {Krebs}\ \emph {et~al.}()\citenamefont {Krebs},
  \citenamefont {Pabst},\ and\ \citenamefont {Santra}}]{KrPa-submitted}%
  \BibitemOpen
  \bibfield  {author} {\bibinfo {author} {\bibfnamefont {D.}~\bibnamefont
  {Krebs}}, \bibinfo {author} {\bibfnamefont {S.}~\bibnamefont {Pabst}}, \ and\
  \bibinfo {author} {\bibfnamefont {R.}~\bibnamefont {Santra}},\
  }\href{\doibase 10.1119/1.4827015} {\enquote {\bibinfo {title} {Introducing
  many-body physics using atomic spectroscopy},}\ }\bibinfo {note} {Am. J.
  Phys., accepted (2013);
  \href{http://arxiv.org/abs/1311.4466v1}{arxiv:1311.4466}}\BibitemShut
  {NoStop}%
\bibitem [{n.p()}]{n.parameters}%
  \BibitemOpen
  \href@noop {} {}\bibinfo {note} {For the radial grid 1000 grid points are
  used with a mapping parameter of $\zeta=1$. The complex absorbing potential
  starts at $220~a_0$ and has a strength of $5\cdot 10^{-3}$. Orbitals with an
  energy larger than 15~a.u. are omitted resulting into around 430 states per
  angular momentum.}\BibitemShut {Stop}%
\bibitem [{\citenamefont {Corkum}(1993)}]{Co-PRL-1993}%
  \BibitemOpen
  \bibfield  {author} {\bibinfo {author} {\bibfnamefont {P.~B.}\ \bibnamefont
  {Corkum}},\ }\href{\doibase 10.1103/PhysRevLett.71.1994} {\bibfield
  {journal} {\bibinfo  {journal} {Phys. Rev. Lett.}\ }\textbf {\bibinfo
  {volume} {71}},\ \bibinfo {pages} {1994} (\bibinfo {year}
  {1993})}\BibitemShut {NoStop}%
\bibitem [{\citenamefont {Schafer}\ \emph {et~al.}(1993)\citenamefont
  {Schafer}, \citenamefont {Yang}, \citenamefont {DiMauro},\ and\ \citenamefont
  {Kulander}}]{ScKu-PRL-1993}%
  \BibitemOpen
  \bibfield  {author} {\bibinfo {author} {\bibfnamefont {K.~J.}\ \bibnamefont
  {Schafer}}, \bibinfo {author} {\bibfnamefont {B.}~\bibnamefont {Yang}},
  \bibinfo {author} {\bibfnamefont {L.~F.}\ \bibnamefont {DiMauro}}, \ and\
  \bibinfo {author} {\bibfnamefont {K.~C.}\ \bibnamefont {Kulander}},\
  }\href{\doibase 10.1103/PhysRevLett.70.1599} {\bibfield  {journal} {\bibinfo
  {journal} {Phys. Rev. Lett.}\ }\textbf {\bibinfo {volume} {70}},\ \bibinfo
  {pages} {1599} (\bibinfo {year} {1993})}\BibitemShut {NoStop}%
\bibitem [{\citenamefont {Lewenstein}\ \emph {et~al.}(1994)\citenamefont
  {Lewenstein}, \citenamefont {Balcou}, \citenamefont {Ivanov}, \citenamefont
  {L'Huillier},\ and\ \citenamefont {Corkum}}]{LeCo-PRA-1994}%
  \BibitemOpen
  \bibfield  {author} {\bibinfo {author} {\bibfnamefont {M.}~\bibnamefont
  {Lewenstein}}, \bibinfo {author} {\bibfnamefont {P.}~\bibnamefont {Balcou}},
  \bibinfo {author} {\bibfnamefont {M.~Y.}\ \bibnamefont {Ivanov}}, \bibinfo
  {author} {\bibfnamefont {A.}~\bibnamefont {L'Huillier}}, \ and\ \bibinfo
  {author} {\bibfnamefont {P.~B.}\ \bibnamefont {Corkum}},\ }\href{\doibase
  10.1103/PhysRevA.49.2117} {\bibfield  {journal} {\bibinfo  {journal} {Phys.
  Rev. A}\ }\textbf {\bibinfo {volume} {49}},\ \bibinfo {pages} {2117}
  (\bibinfo {year} {1994})}\BibitemShut {NoStop}%
\bibitem [{n.s()}]{n.spin}%
  \BibitemOpen
  \href@noop {} {}\bibinfo {note} {Excluding the spin degree of
  freedom}\BibitemShut {NoStop}%
\bibitem [{n.v()}]{n.velocity}%
  \BibitemOpen
  \href@noop {} {}\bibinfo {note} {The HHG spectra and the photoionization
  cross sections are calcualted using the velocity form of the dipole
  transition~\cite{Pa-EPJST-2013}.}\BibitemShut {Stop}%
\bibitem [{\citenamefont {Morishita}\ \emph {et~al.}(2008)\citenamefont
  {Morishita}, \citenamefont {Le}, \citenamefont {Chen},\ and\ \citenamefont
  {Lin}}]{MoLi-PRL-2008}%
  \BibitemOpen
  \bibfield  {author} {\bibinfo {author} {\bibfnamefont {T.}~\bibnamefont
  {Morishita}}, \bibinfo {author} {\bibfnamefont {A.-T.}\ \bibnamefont {Le}},
  \bibinfo {author} {\bibfnamefont {Z.}~\bibnamefont {Chen}}, \ and\ \bibinfo
  {author} {\bibfnamefont {C.~D.}\ \bibnamefont {Lin}},\ }\href{\doibase
  10.1103/PhysRevLett.100.013903} {\bibfield  {journal} {\bibinfo  {journal}
  {Phys. Rev. Lett.}\ }\textbf {\bibinfo {volume} {100}},\ \bibinfo {pages}
  {013903} (\bibinfo {year} {2008})}\BibitemShut {NoStop}%
\bibitem [{\citenamefont {Le}\ \emph {et~al.}(2009)\citenamefont {Le},
  \citenamefont {Lucchese}, \citenamefont {Tonzani}, \citenamefont
  {Morishita},\ and\ \citenamefont {Lin}}]{LeLi-PRA-2009}%
  \BibitemOpen
  \bibfield  {author} {\bibinfo {author} {\bibfnamefont {A.-T.}\ \bibnamefont
  {Le}}, \bibinfo {author} {\bibfnamefont {R.~R.}\ \bibnamefont {Lucchese}},
  \bibinfo {author} {\bibfnamefont {S.}~\bibnamefont {Tonzani}}, \bibinfo
  {author} {\bibfnamefont {T.}~\bibnamefont {Morishita}}, \ and\ \bibinfo
  {author} {\bibfnamefont {C.~D.}\ \bibnamefont {Lin}},\ }\href{\doibase
  10.1103/PhysRevA.80.013401} {\bibfield  {journal} {\bibinfo  {journal} {Phys.
  Rev. A}\ }\textbf {\bibinfo {volume} {80}},\ \bibinfo {pages} {013401}
  (\bibinfo {year} {2009})}\BibitemShut {NoStop}%
\bibitem [{\citenamefont {Starace}(1982)}]{St-Springer-1980}%
  \BibitemOpen
  \bibfield  {author} {\bibinfo {author} {\bibfnamefont {A.~F.}\ \bibnamefont
  {Starace}},\ }in\
  \href{http://www.springer.com/materials/book/978-3-642-46455-3} {\emph
  {\bibinfo {booktitle} {Encyclopedia of Physics}}},\ Vol.\ \bibinfo {volume}
  {31: Corpuscles and Radiation in Matter I},\ \bibinfo {editor} {edited by\
  \bibinfo {editor} {\bibfnamefont {W.}~\bibnamefont {Mehlhorn}}}\ (\bibinfo
  {publisher} {Springer, Berlin},\ \bibinfo {year} {1982})\ Chap.\ \bibinfo
  {chapter} {Theory of Atomic Photoionization}, pp.\ \bibinfo {pages}
  {1--121}\BibitemShut {NoStop}%
\bibitem [{n.r()}]{n.radius}%
  \BibitemOpen
  \href@noop {} {}\bibinfo {note} {These estimations are based on the radial
  moments $\left<\varphi_i|r^n|\varphi_i\right>$ of the occupied Hartree-Fock
  orbitals $\varphi_i$.}\BibitemShut {Stop}%
\bibitem [{n.e()}]{n.espec}%
  \BibitemOpen
  \href@noop {} {}\bibinfo {note} {By using this formula, we are able to
  compare directly with the statements made in
  Ref.~\cite{ShVi-NatPhys-2011}.}\BibitemShut {Stop}%
\bibitem [{\citenamefont {Boguslavskiy}\ \emph {et~al.}(2012)\citenamefont
  {Boguslavskiy}, \citenamefont {Mikosch}, \citenamefont {Gijsbertsen},
  \citenamefont {Spanner}, \citenamefont {Patchkovskii}, \citenamefont {Gador},
  \citenamefont {Vrakking},\ and\ \citenamefont {Stolow}}]{BoMi-Science-2012}%
  \BibitemOpen
  \bibfield  {author} {\bibinfo {author} {\bibfnamefont {A.~E.}\ \bibnamefont
  {Boguslavskiy}}, \bibinfo {author} {\bibfnamefont {J.}~\bibnamefont
  {Mikosch}}, \bibinfo {author} {\bibfnamefont {A.}~\bibnamefont
  {Gijsbertsen}},  \emph {et~al.},\ }\href{\doibase 10.1126/science.1212896}
  {\bibfield  {journal} {\bibinfo  {journal} {Science}\ }\textbf {\bibinfo
  {volume} {335}},\ \bibinfo {pages} {1336} (\bibinfo {year}
  {2012})}\BibitemShut {NoStop}%
\bibitem [{\citenamefont {Patchkovskii}\ \emph {et~al.}(2006)\citenamefont
  {Patchkovskii}, \citenamefont {Zhao}, \citenamefont {Brabec},\ and\
  \citenamefont {Villeneuve}}]{PaVi-PRL-2006}%
  \BibitemOpen
  \bibfield  {author} {\bibinfo {author} {\bibfnamefont {S.}~\bibnamefont
  {Patchkovskii}}, \bibinfo {author} {\bibfnamefont {Z.}~\bibnamefont {Zhao}},
  \bibinfo {author} {\bibfnamefont {T.}~\bibnamefont {Brabec}}, \ and\ \bibinfo
  {author} {\bibfnamefont {D.~M.}\ \bibnamefont {Villeneuve}},\ }\href{\doibase
  10.1103/PhysRevLett.97.123003} {\bibfield  {journal} {\bibinfo  {journal}
  {Phys. Rev. Lett.}\ }\textbf {\bibinfo {volume} {97}},\ \bibinfo {pages}
  {123003} (\bibinfo {year} {2006})}\BibitemShut {NoStop}%
\bibitem [{\citenamefont {Vozzi}\ \emph
  {et~al.}(2011{\natexlab{b}})\citenamefont {Vozzi}, \citenamefont {Negro},
  \citenamefont {Calegari}, \citenamefont {Sansone}, \citenamefont {Nisoli},
  \citenamefont {De~Silvestri},\ and\ \citenamefont
  {Stagira}}]{VoNe-NatPhys-2011}%
  \BibitemOpen
  \bibfield  {author} {\bibinfo {author} {\bibfnamefont {C.}~\bibnamefont
  {Vozzi}}, \bibinfo {author} {\bibfnamefont {M.}~\bibnamefont {Negro}},
  \bibinfo {author} {\bibfnamefont {F.}~\bibnamefont {Calegari}}, \bibinfo
  {author} {\bibfnamefont {G.}~\bibnamefont {Sansone}}, \bibinfo {author}
  {\bibfnamefont {M.}~\bibnamefont {Nisoli}}, \bibinfo {author} {\bibfnamefont
  {S.}~\bibnamefont {De~Silvestri}}, \ and\ \bibinfo {author} {\bibfnamefont
  {S.}~\bibnamefont {Stagira}},\ }\href{http://dx.doi.org/10.1038/nphys2029}
  {\bibfield  {journal} {\bibinfo  {journal} {Nat. Phys.}\ }\textbf {\bibinfo
  {volume} {7}},\ \bibinfo {pages} {822} (\bibinfo {year}
  {2011}{\natexlab{b}})}\BibitemShut {NoStop}%
\bibitem [{\citenamefont {Awasthi}\ \emph {et~al.}(2005)\citenamefont
  {Awasthi}, \citenamefont {Vanne},\ and\ \citenamefont
  {Saenz}}]{AwVa-JPB-2005}%
  \BibitemOpen
  \bibfield  {author} {\bibinfo {author} {\bibfnamefont {M.}~\bibnamefont
  {Awasthi}}, \bibinfo {author} {\bibfnamefont {Y.~V.}\ \bibnamefont {Vanne}},
  \ and\ \bibinfo {author} {\bibfnamefont {A.}~\bibnamefont {Saenz}},\
  }\href{http://stacks.iop.org/0953-4075/38/i=22/a=005} {\bibfield  {journal}
  {\bibinfo  {journal} {J. Phys. B: At., Mol. Opt. Phys.}\ }\textbf {\bibinfo
  {volume} {38}},\ \bibinfo {pages} {3973} (\bibinfo {year}
  {2005})}\BibitemShut {NoStop}%
\bibitem [{\citenamefont {Schlegel}\ \emph {et~al.}(2007)\citenamefont
  {Schlegel}, \citenamefont {Smith},\ and\ \citenamefont
  {Li}}]{ScSm-JCP126-2007}%
  \BibitemOpen
  \bibfield  {author} {\bibinfo {author} {\bibfnamefont {H.~B.}\ \bibnamefont
  {Schlegel}}, \bibinfo {author} {\bibfnamefont {S.~M.}\ \bibnamefont {Smith}},
  \ and\ \bibinfo {author} {\bibfnamefont {X.}~\bibnamefont {Li}},\
  }\href{\doibase 10.1063/1.2743982} {\bibfield  {journal} {\bibinfo  {journal}
  {J. Chem. Phys.}\ }\textbf {\bibinfo {volume} {126}},\ \bibinfo {eid}
  {244110} (\bibinfo {year} {2007})}\BibitemShut {NoStop}%
\bibitem [{\citenamefont {Klinkusch}\ \emph {et~al.}(2009)\citenamefont
  {Klinkusch}, \citenamefont {Saalfrank},\ and\ \citenamefont
  {Klamroth}}]{KlSa-JCP131-2009}%
  \BibitemOpen
  \bibfield  {author} {\bibinfo {author} {\bibfnamefont {S.}~\bibnamefont
  {Klinkusch}}, \bibinfo {author} {\bibfnamefont {P.}~\bibnamefont
  {Saalfrank}}, \ and\ \bibinfo {author} {\bibfnamefont {T.}~\bibnamefont
  {Klamroth}},\ }\href{\doibase 10.1063/1.3218847} {\bibfield  {journal}
  {\bibinfo  {journal} {J. Chem. Phys.}\ }\textbf {\bibinfo {volume} {131}},\
  \bibinfo {eid} {114304} (\bibinfo {year} {2009})}\BibitemShut {NoStop}%
\bibitem [{\citenamefont {Spanner}\ and\ \citenamefont
  {Patchkovskii}(2009)}]{SpPa-PRA-2009}%
  \BibitemOpen
  \bibfield  {author} {\bibinfo {author} {\bibfnamefont {M.}~\bibnamefont
  {Spanner}}\ and\ \bibinfo {author} {\bibfnamefont {S.}~\bibnamefont
  {Patchkovskii}},\ }\href{\doibase 10.1103/PhysRevA.80.063411} {\bibfield
  {journal} {\bibinfo  {journal} {Phys. Rev. A}\ }\textbf {\bibinfo {volume}
  {80}},\ \bibinfo {pages} {063411} (\bibinfo {year} {2009})}\BibitemShut
  {NoStop}%
\bibitem [{\citenamefont {Torlina}\ \emph {et~al.}(2012)\citenamefont
  {Torlina}, \citenamefont {Ivanov}, \citenamefont {Walters},\ and\
  \citenamefont {Smirnova}}]{ToIv-RPA-2012}%
  \BibitemOpen
  \bibfield  {author} {\bibinfo {author} {\bibfnamefont {L.}~\bibnamefont
  {Torlina}}, \bibinfo {author} {\bibfnamefont {M.}~\bibnamefont {Ivanov}},
  \bibinfo {author} {\bibfnamefont {Z.~B.}\ \bibnamefont {Walters}}, \ and\
  \bibinfo {author} {\bibfnamefont {O.}~\bibnamefont {Smirnova}},\
  }\href{\doibase 10.1103/PhysRevA.86.043409} {\bibfield  {journal} {\bibinfo
  {journal} {Phys. Rev. A}\ }\textbf {\bibinfo {volume} {86}},\ \bibinfo
  {pages} {043409} (\bibinfo {year} {2012})}\BibitemShut {NoStop}%
\end{thebibliography}%


\end{document}